# Further investigations on the interface instability between fresh injections and burnt products in 2-D rotating detonation

Qin Li [a,b], Pengxin Liu [a], Hanxin Zhang [a,b]

[a] State Key Laboratory of Aerodynamics, Mianyang, Sichuan, 621000, China,
and
[b] National Laboratory of Computational Fluid Dynamics, Beijing University of Aeronautics and Astronautics, Beijing, 100191, China

**Abstract** Instabilities in rotating detonation are concerned because of their potential influence on the stability of operation. Previous studies on instability of 2-D rotating detonation mainly cared about the one of the contact discontinuity originated from the conjunction of the detonation and oblique shock. Hishida et al. first found the rippled structure existed in the interface between fresh injections and burnt product from the previous cycle (Shock Waves 19, 2009:1–10), and a mechanism of Kelvin-Helmholtz instability was suggested as well. Similar structures were observed as well in simulations by current authors, where a fifth-order WENO-type scheme with improved resolution and 7-species-and-8-reaction chemical model on $H_2$/air mixture were used for solving Euler equations. In order to achieve a deep understanding on the flow mechanism, more careful simulations are carried out by using three grids with increasing resolution. The results show that besides the previously-mentioned Kelvin-Helmholtz instability, there are two other mechanisms which take effect in the interface instability, i.e., the effect of baroclinic torque and Rayleigh-Taylor instability. Occurrence conditions for two instabilities are checked and testified. Especially, the spike- and bubble-like structures are observed at the interface, which show appearances different from canonical structures by Kelvin-Helmholtz instability.

**1. Introduction.**

Considered as the isochoric-like combustion, detonation is thought to have faster heat release, less entropy increase and higher thermodynamic efficiency compared with the deflagration [1]. It is well-known that there are two main-streams of facilities regarding detonation, i.e., pulsed and rotating detonation engine (abbreviated as PDE and RDE respectively). Recently, it is mentioned in Ref. [2] that a global shift of research emphasis has occurred from PDE to RDE.

Although the feasibility of RDE was studied sixty years ago by Voitsekhovskii, Nicholls, and etc. [3], substantial advances seemed to be achieved since 1980s [4]. Later on, investigations were populated in experiments and simulations [1, 3-6]. Comprehensive topics were concerned such as injection techniques, ignitions, the geometry of combustors and exhaust nozzles, the mechanism of rotating detonation, the chemical dynamics of reactions, parametric window of operations, the performance and thermo-cycle analysis, so on and so forth.

Because in experiments rotating detonation usually runs inside a solid annular chamber at the speed of kilometers per second, detailed diagnostic is difficult and explicit information of flow structure is usually limited. In this regard, numerical simulation can serve as a powerful tool to reconstruct the flow field with enough details. Furthermore, parametric studies are observed by simulations to explore existence domains and operation boundaries [7-9]. Despite the advantages, it should be mentioned that large simplifications are usually applied in simulations, e.g., the use of 2-D model by unwrapping 3-D annular configuration, the pre-mixture assumption of reactants, simplified chemical model, neglecting of injection details and the absence of viscosity in many studies. Even so, simulations are expected to reveal the main mechanism in rotating detonation. With the advancement of computational fluid dynamics, more details start to be considered in simulations, e.g., the injection process though nozzles was simulated in Ref. [10] and the viscosity was considered in Refs. [11-12].

From literatures, the basic structures of rotating detonation is known as [4]: before the detonation wave, there is a triangular zone of fresh combustibles, where pre-mixtures are kept feeding and increase the height until meeting the detonation; within the detonation, isochoric-like reaction happens and reactants are consumed; after the detonation, a region with high pressure appears followed by rarefactions; at the longitudinal end of the detonation, an oblique shock wave is formed and directs axial downstream due to the lateral expansion of products; at the conjunction of the shock wave with the detonation, a contact discontinuity is generated and usually induces flow instability. The circumferential downstream of detonation is connected to the beginning of reactants injection due to the annular circulation, and sometimes multiple detonations might occur depending on situations. With the advances of simulations, more details are discovered continuously [1, 3, 6, 8, 12-14].

The stable operation of RDE is one of targets pursued by investigators. There are many factors influencing the operation, and one of which might be the flow instability. The instability problem is also interested by fundamental researches. As mentioned before, the contact discontinuity is widely reported by literatures, which is apt to cause instability and generate vortices. In Ref. [3], it is stated that the contact surface which lays between the newly injected propellants and the previously burnt products should be concerned, which might destabilize the detonation wave by reducing its height or degenerating the detonation into deflagration. In Ref. [14], Hishida et al. first reported the existence of rippled structures at aforementioned interface of a 2-D simulation. They conjectured that the structure might arise from Kelvin-Helmholtz (K-H) instability. To confirm this, the convective velocity $U_c$ was predicted and compared with the one derived from the simulation. A fair agreement was obtained. However, the structure of ripples is not distinct enough and it is not known if the K-H instability should be the only mechanism.

In order to further explore the uncertainties, numerical simulations are carried out by using a fifth-order WENO-type scheme [15] and 7-species-and-8-reaction chemical model. In Ref. [15], the scheme has manifested its high resolution to resolve subtle flow structures along slip line, and the capability is just needed by current investigation. In Section 2, numerical methods and their validations are first introduced. In Section 3, results of simulation are generally introduced. In Section 4, instabilities at the interface between fresh injections and burnt products are analyzed in detail, where two other mechanisms are proposed besides the previous Kelvin-Helmholtz instability. At last, the conclusion is drawn in Section 5.

**2. Numerical methods.**

2.1. Governing equations.

Although the viscosity takes effect as mentioned in Ref. [11], the inviscid Euler equations with chemical reactions have been widely employed because of the much less computational cost. Hence the non-dimensional Euler equations are chosen in this study with *ns*-species-and-*nr*-reaction model considered. On consideration of applicability, the 3-D equations are chosen to solve 2-D problem as:

$$\frac{\partial Q}{\partial t} + \frac{\partial E}{\partial x} + \frac{\partial F}{\partial y} + \frac{\partial G}{\partial z} = S, \qquad (1)$$

where

$$Q = \begin{pmatrix} \rho \\ \rho u \\ \rho v \\ \rho w \\ E \\ \rho f_i \end{pmatrix}, \ E = \begin{pmatrix} \rho u \\ \rho u^2 + P \\ \rho uv \\ \rho uw \\ (E+p)u \\ \rho u f_i \end{pmatrix}, \ F = \begin{pmatrix} \rho v \\ \rho vu \\ \rho v^2 + p \\ \rho vw \\ (E+p)v \\ \rho v f_i \end{pmatrix}, \ G = \begin{pmatrix} \rho w \\ \rho wu \\ \rho wv \\ \rho w^2 + p \\ (E+p)w \\ \rho w f_i \end{pmatrix}, \ S = \begin{pmatrix} 0 \\ 0 \\ 0 \\ 0 \\ 0 \\ S_i \end{pmatrix}, \ f_i \text{ is the}$$

mass fraction, $S_i$ is the chemical source term, and where the subscript "*i*" corresponds to the species *i*. The total energy $E$ in $Q$ is defined as

$$E = \rho \sum_i f_i h_i - p + \frac{1}{2}\rho(u^2 + v^2 + w^2), \qquad (2)$$

where $h_i$ is the specific enthalpy. The non-dimensionalization is carried out and exemplified as

$$x = \frac{x^*}{L_\infty}, u = \frac{u^*}{V_\infty}, \rho = \frac{\rho^*}{\rho_\infty}, p = \frac{p^*}{\rho_\infty V_\infty^2}, T = \frac{T^*}{T_\infty}, E = \frac{E^*}{\rho_\infty V_\infty^2}, h = \frac{h^*}{V_\infty^2}, S_i = S_i^* \frac{L_\infty}{\rho_\infty V_\infty}, \qquad (3)$$

where the superscript '*' denotes the dimensional state of variables. In Eq. (2), $h_i$ is derived from its dimensional counterpart by using polynomial fitting of the temperature such as

$$h_i^*(T^*) = \frac{R_0}{M_i^*} \sum_{j=0}^{5} A_{i,j}(T^*)^j, \qquad (4)$$

where $A_{i,j}$ are coefficients with respect to species and are exemplified in the Appendix. The dimensional generation source term $S_i^*$ is derived from the following reversible reaction model:

$$\sum_{k=1}^{ns} \alpha_{jk} X_k \Leftrightarrow \sum_{k=1}^{ns} \beta_{jk} X_k, \ j=1,...,nr, \qquad (5)$$

where $X_k$ denotes the species in reactions such as $H_2$, $\alpha_{jk}$ and $\beta_{jk}$ are coefficients in the reaction formula, *ns* is the number of species, and *nr* is the number of reactions. Then $S_i^*$ can be given as

$$S_i = M_i^* \sum_{j=1}^{nr} (\beta_{ji} - \alpha_{ji})(R_j - R_{-j}), \tag{6}$$

where $M_i^*$ is the molecular weight of the *i*-th species, and $R_j^*$ and $R_{-j}^*$ are the speed rate of forward and reverse reactions. The rates are computed by

$$\begin{cases} R_j = K_j \prod_{i=1}^{ns} \left(\frac{\rho_i}{M_i}\right)^{\alpha_{ji}} \left(\sum_{k=1}^{ns} \frac{\rho_k}{M_k} C_{jk}\right)^{L_j} \\ R_{-j} = K_{-j} \prod_{i=1}^{ns} \left(\frac{\rho_i}{M_i}\right)^{\beta_{ji}} \left(\sum_{k=1}^{ns} \frac{\rho_k}{M_k} C_{jk}\right)^{L_j} \end{cases}, \tag{7}$$

where $K_j^*$ and $K_{-j}^*$ are constant coefficient of reaction rates, $C_{jk}$ is the collision efficiency with the value one in current computation, and where $L_i$ is a switch parameter to identify the triple-object collision, namely, $L_i = 1$ indicates the collision occurrence and otherwise $L_i = 0$. $K_j^*$ and $K_{-j}^*$ are calculated by Arrhenius law as

$$K_j^* = A_j (T^*)^{B_j} e^{(-C_j/T^*)} \quad \text{and} \quad K_{-j}^* = A_{-j}(T^*)^{B_{-j}} e^{(-C_{-j}/T^*)}, \tag{8}$$

where $A_j$, $B_j$, $C_j$ and similar counterparts are species-related coefficients. Through Eqs. (1)-(8), the general *ns*-species-and-*nr*-reaction model can be implemented with the integration with aerodynamic equations. In this study, a specific 7-species-and-8-reaction one is chosen and corresponding parameters can be found in the Appendix.

On consideration of applicability, Eq. (1) is solved in the general computational coordinate system by the grid transformation: $(x, y, z) \to (\xi, \eta, \zeta)$ as

$$\frac{\partial \hat{Q}}{\partial t} + \frac{\partial \hat{E}}{\partial \xi} + \frac{\partial F}{\partial \eta} + \frac{\partial G}{\partial \zeta} = \hat{S}, \tag{9}$$

where $\hat{Q} = J^{-1} Q$, $\hat{S} = J^{-1} S$, $\hat{E} = (\hat{\xi}_x E + \hat{\xi}_y F + \hat{\xi}_z G)$, $J^{-1} = \left|\frac{\partial(x,y,z)}{\partial(\xi,\eta,\zeta)}\right|$, $(\hat{\xi}_x, \hat{\xi}_y, \hat{\xi}_z) = J^{-1}(\vec{r}_\eta \times \vec{r}_\zeta)$, and $\hat{F}, \hat{G}$ can be obtained similarly.

2.2. Numerical schemes.

In order to achieve a simulation with high resolution, the numerical scheme to discretize the derivatives in Eq. (9) is of critical importance. An improved fifth-order WENO [15] is used in this study for spatial discretization, and a brief review is given as following.

Taking the 1-D hyperbolic conservative law as an example,

$$u_t + f(u)_x = 0. \tag{10}$$

Suppose the grids are equally partitioned as $x_j = j\Delta x$ where $\Delta x$ denotes the interval and *j* is the grid index, Eq. (10) at $x_j$ can be re-written in conservative form as

$$(u_t)_j = -\left(\hat{f}_{j+1/2} - \hat{f}_{j-1/2}\right)/\Delta x, \tag{11}$$

where $\hat{f}_{j+1/2}$ is the evaluation of $\hat{f}(x)$ at $x_{j+1/2}$, and $\hat{f}(x)$ is implicitly defined by

$f(x) = \frac{1}{\Delta x}\int_{x-\Delta x/2}^{x+\Delta x/2}\hat{f}(x')dx'$. Usually, the flux $\hat{f}(x)$ is split into $\hat{f}^+$ and $\hat{f}^-$ according to eigenvalues of $\partial \hat{f}(u)/\partial u$. Taking $f^+$ as an example and dropping the superscript '+' for brevity, the standard WENO scheme can be formulated as [16]:

$$\hat{f}_{j+1/2} = \sum_{k=0}^{r}\omega_k q_k^r, \tag{12}$$

where $r$ represents the grid-stencil number (e.g., 3 for the fifth-order WENO5), $\omega_k$ is the nonlinear weight derived from the linear counterpart $C_k^r$, and $q_k^r$ denotes the candidate scheme on basic stencils as $q_k^r = \sum_{l=0}^{r-1}a_{k,l}^r f(u_{j-r+k+l+1})$. For WENO5, $r=3$, $C_k^3 = \{0.3, 0.6, 0.1\}$ and $a_{k,l}^3$ can be found in Ref. [16] accordingly. For brevity, the superscript '$r$' is dropped afterwards. The derivation of $\omega_k$ is by

$$\omega_k = \alpha_k \Big/ \sum_{l=0}^{2}\alpha_l, \tag{13}$$

where

$$\alpha_k = C_k \Big/ (\varepsilon + IS_k)^p \tag{14}$$

and usually $p=2$ and $\varepsilon=10^{-5} \sim 10^{-7}$ in Ref. [16]. In Eq. (14), $IS_k$ is the smoothness indicator, and ones in Ref. [16] is defined as (called as $IS^{JS}$)

$$IS_k^{JS} = \sum_{l=1}^{2}\int_{x_{j-1/2}}^{x_{j+1/2}}\Delta x^{2l-1}\left(\partial^{(l)}q_k(x)/\partial x^{(l)}\right)^2 dx., \tag{15}$$

where $q_k(x)$ is the reconstruction polynomial. For WENO5,

$$\begin{cases} IS_0^{JS} = \frac{13}{12}(f_{j-2} - 2f_{j-1} + f_j)^2 + \frac{1}{4}(f_{j-2} - 4f_{j-1} + 3f_j)^2 \\ IS_1^{JS} = \frac{13}{12}(f_{j-1} - 2f_j + f_{j+1})^2 + \frac{1}{4}(f_{j-1} - f_{j+1})^2 \\ IS_2^{JS} = \frac{13}{12}(f_j - 2f_{j+1} + f_{j+2})^2 + \frac{1}{4}(3f_j - 4f_{j+1} + f_{j+2})^2 \end{cases}. \tag{16}$$

In Ref. [15], new piecewise-polynomial mapping functions $g_k$ were proposed to improve the performance of WENO5, which is fulfilled by invoking a revision on $\omega_k$ derived by Eq. (13). A fifth-order version of mapping functions is used here as

$$\omega_k' = g_k(\omega_k) = \begin{cases} C_k\left[1 + (a-1)^5\right] & \omega_k \leq C_k \\ C_k + b^4(\omega - C_k)^5 & \omega_k > C_k \end{cases}, \tag{17}$$

where $a = \omega_k/C_k$ and $b = 1/(C_k - 1)$. At last, the final nonlinear weights $\omega_k$ will be acquired by normalizing the newly obtained $\omega_k'$ through $\omega_k = \omega_k'/\sum_{l=0}^{2}\omega_l'$. The corresponding scheme is called as WENO-PPM5. In Ref. [15], the prominent capability of the scheme had been manifested on resolving subtle structures along the contact surface, which should be suitable to investigate instabilities in current study. More details are suggested to Ref. [15].

For unsteady problems, the evaluation of temporal derivative is quite important. There are at least two approaches for approximations, i.e., the multi-step Runge-Kutta method and the second-order dual time-step method. The former can achieve series accurate orders and the latter is thought to be more stable to handle the "stiff" problem in reaction flows. Currently, it is found that a third-order Total-variation-diminishing (TVD) Runge-Kutta scheme [16], which is widely used in computational fluid dynamics, works well in problems investigated. Some validating cases will be shown in Section 2.4.

2.3. Boundary conditions.

There are three types of boundaries in 2-D flows, i.e., the headwall for injection, exhaust boundary and lateral boundary. The last one is originated from the circumferential configuration of the 3-D annular chamber, and apparently the periodic boundary condition should be used. For the headwall boundary, most of simulations have chosen simplified models other than directly simulating the complexity of the real injection. The simplifications usually lie in two aspects, namely, the pre-mixture assumption of reactants and numerical boundary conditions to simulate unsteady injection [7- 9, 12, 14]. In this study, the condition similar to Ref. [7] is chosen where three situations are considered. Let $P_w$ be the pressure extrapolated from the inner field to the headwall, then variables at the head wall ($P$, $T$, the normal ($u$) and tangential ($v$) velocity with respect to the wall) are derived according to the following situations:

(1) Blocked state: If $P_w > P_0$ where $P_0$ is the total pressure of the injection, then $u = 0$; $P$, $\rho$ and $v$ will be extrapolated from the inner field.

(2) Subsonic injection: If $P_0 > P_w > P_{cr}$ where $P_{cr} = P_0 \left(\frac{2}{\gamma+1}\right)^{\frac{\gamma}{\gamma-1}}$, $P = P_w$, $v = 0$ and $u$ is derived by

$$u = \sqrt{\frac{2\gamma}{\gamma-1}\left[1 - \left(\frac{P}{P_0}\right)^{\frac{\gamma-1}{\gamma}}\right]}, \quad (18)$$

and $T$ is updated by:

$$T = T_0 \left(\frac{P}{P_0}\right)^{\frac{\gamma-1}{\gamma}}, \quad (19)$$

where $T_0$ is the total temperature and $\gamma$ is the specific heat ratio of the mixture. $\gamma$ is derived by

$$\gamma = \frac{\sum_{i=1}^{ns}(f_i C_{p,i})}{\sum_{i=1}^{ns}(f_i C_{p,i}) - R}, \quad (20)$$

$$C_{p,i} = \frac{R_0}{M_i^*}\sum_{j=1}^{5}\left(A_{i,j}^{C_p}(T^*)^{(j-1)}\right), \quad (21)$$

$$R = R_0 \sum_{i=1}^{ns}\left(\frac{f_i}{M_i^*}\right), \quad (22)$$

where $R_0 = 8.31434 \text{J}/(\text{Mol} \times \text{K})$ and values of coefficients $A_{i,j}^{C_p}$ are shown in the Appendix.

(3) Sonic injection: If $P_{cr} > P_w$, $P = P_{cr}$ and $T$ is derived as in (1); $v = 0$ and $u$ is derived by

$$u = \sqrt{\frac{2\gamma}{\gamma+1}RT_0} \tag{23}$$

For exhaust boundary, the condition depends on the local Mach number $M$: if $M > 1$, all variables at the boundary are extrapolated from the inner field; if $M \leq 1$, similar procedure is implemented except $P$ is set as the ambient pressure.

2.4. Validating tests.

To verify the computational framework, two cases are tested: the propagation of 1-D detonation and the combustion induced by the shock around a sphere. The governing equations are Euler equations with the gas mixtures $H_2$/Air considered, and a 7-species-and-8-reaction chemical model by Evans & Schexnayder [18] is adopted. The detail of the model can be found in the Appendix. In computations, WENO5-PPM5 is used for discretization and the Steger-Warming scheme is used for flux splitting.

(1) Propagation of 1-D detonation.

The stoichiometric reactants in a 30cm-long tube is considered. The initial conditions in most part of the tube are: $P = 2atm$ and $T = 400K$. To trigger the detonation, an artificial ignition is imposed at a small region [0, 0.5cm] by setting $P = 30atm$ and $T = 6000K$. For comparison, the theoretical Chapman-Jouguet (C-J) properties are provided as [7]: the detonation velocity ($U_D$) is 1979.1m/s, C-J pressure is 23.64atm and C-J temperature is 3028.9K. The third-order TVD Runge-Kutta method is used for temporal discretization. Ref. [7] indicated that the accuracy of simulations might influence the propagation speed of discontinuities.

Three grid sizes are checked, namely, $\Delta x$=0.5mm, 0.3mm and 0.1mm. The distribution of pressure and temperature with dimension are depicted in Fig. 1 at $t$=120.0$\mu s$ under $\Delta t$=10$^{-3}\mu s$. From the figure, curves of the pressure and temperature distributions at different $\Delta x$ almost fall together respectively, which indicates the convergence of results. In Table 1, comparisons between the computation and theoretical values are given. From the results, it can be seen that the grid convergence is attained and an acceptable agreement is observed. Hence, the capability of predication on unsteady detonation is indicated.

Table 1. Computed C-J properties and corresponding errors of 1-D detonation

|  | $\Delta x$(mm) | $U_D$ (m/s) | Error | $P_{CJ}$(atm) | Error | $T_{CJ}$(K) | Error |
| --- | --- | --- | --- | --- | --- | --- | --- |
| Theoret. Val. | -- | 1979.1 | -- | 23.64 | -- | 3028.9 | -- |
| Computation | 0.1 | 1978.0 | 0.06% | 22.81 | 3.5% | 3024.5 | 0.15% |
|  | 0.3 | 1976.4 | 0.14% | 22.78 | 3.6% | 3023.5 | 0.18% |
|  | 0.5 | 1975.3 | 0.19% | 22.75 | 3.8% | 3022.9 | 0.20% |

(2) Combustion induced by the shock around a sphere at $Ma$=3.5612.

This case is widely used to test the fidelity of simulations on chemical non-equilibrium flow, and experimental data is available for comparison by Lehr [17]. The inflow conditions are: the free-stream velocity is 1892 m/s, temperature is 291 K and the pressure is 24797 Pa. The sphere boundary is treated as an inviscid non-penetration one, where non-catalytic condition is imposed as $\partial f_i/\partial n = 0$ with $n$ denoting the normal direction. The grid number is:

$n_{\text{streamwise}} \times n_{\text{normal}} \times n_{\text{circumferential}} = 63 \times 71 \times 31$. Because of the steady nature, LU-SGS is chosen for temporal approximation.

In Fig. 2, the temperature contours show a well agreement of predicted shock profile with that of the experiment. In Fig. 3, distributions of pressure and temperature are drawn along the stagnation line, while mass fraction distributions of various species are shown in Fig. 4. From the figures, good quantitative agreement is obtained between the computation and experiment, which indicate the solver is qualified to solve flows with chemical reaction.

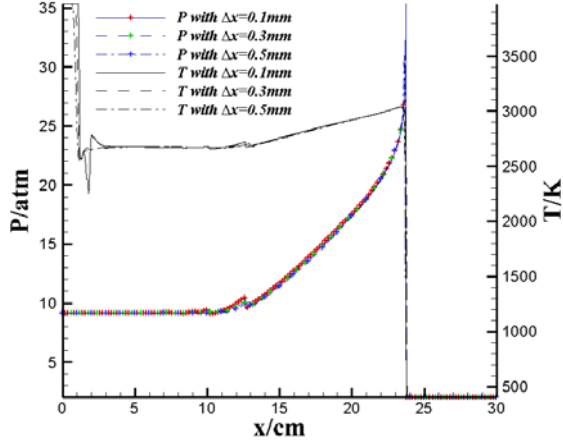

Fig. 1. Distributions of pressure (line-cross) and temperature (line) with dimension at $t=120.0\mu s$ under different $\Delta x$.

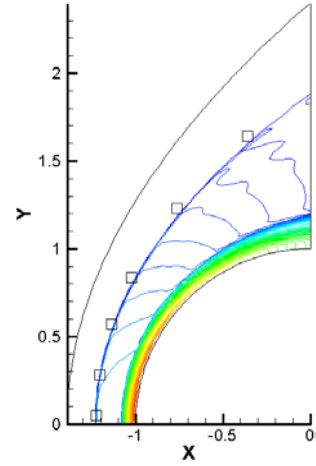

Fig. 2. Contours of non-dimensional temperature of combustion induced by shock. (Lines: 53 levels from 1.8 to 11.4; Square: experiment)

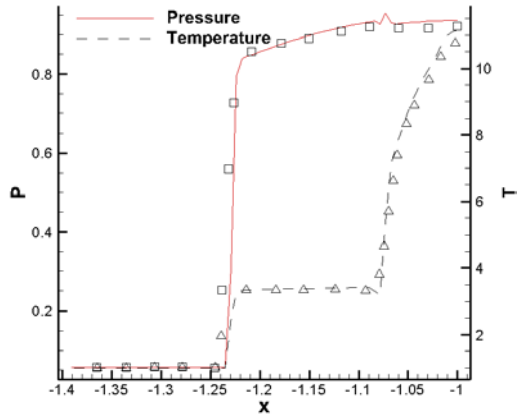

Fig. 3. Distributions of non-dimensional pressure and temperature along stagnation line. (Lines: computation; Symbols: experiment)

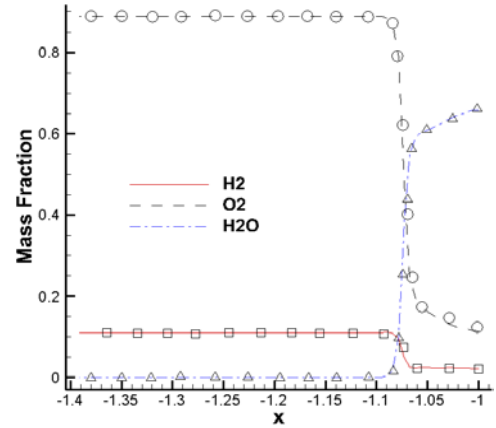

Fig. 4. Distributions of mass fraction of various components along stagnation line. (Lines: computation; Symbols: experiment)

**3. Results.**

Using the methods discussed in Section 2, simulations are made on a 2-D rotating detonation. The case setup and results of the computation are introduced first.

3.1. Case setup.

A rectangular domain is used for 2-D simulation as the approximation of an unwrapping of 3-D circular chamber. In 2-D situation, *x*- and *y*-axis correspond to the axial and circumferential directions in 3-D counterpart respectively. The range of the domain in *x* and *y* direction is: 40 mm×100 mm. Mixtures of $H_2$/Air are used as reactants injected at the head wall where *x*=0, and the stoichiometric ratio is selected for reactants or the Mole ratio for $H_2$, $O_2$, and $N_2$ is 2:1:3.76. The stagnation parameters of the injection and the ambient pressure $P_e$ of the exhaustion are: $P_0$=0.35MPa, $T_0$=300K, and $P_e$=0.1MPa.

In order to investigate the effect of grid size, three sets of grids are used with increasing number ($n_x \times n_y$) as: 201×501 for Case 1, 401×1001 for Case 2 and 801×2001 for Case 3. In order to elucidate why grids with different resolutions are chosen, the similar situation in studies on boundary-layer instability is used for explanation. It is well known that unstable structures there cannot be reproduced by simulations when coarse grids are used, but corresponding grids are usually qualified to describe the steady parametric profile accurately for instability analysis. If keeping refining grids, unstable structures in boundary layers can be resolved finally when the grid length-scale is small enough. Similar idea is employed in current study as shown in Section 4: relatively coarse grids are first used in Case 1 and quasi-stable flow field is obtained which serves as the base one for instability analysis; with the refinement of grids in Case 2, ripples occur at the reactant interface, which reproduces the observation in Ref. [14] and is also regarded as a preliminary show of the instability; when grids are further refined in Case 3, ripples are replaced by well-organized vortex-like subtleties, and mechanisms especially extra ones are planned to discuss in Section 4.

A numerical ignition is used to induce the detonation as follows. First, the upper region at 20mm<*y*<100mm is filled with combustibles and the rest field is filled with air. The initial pressure in the whole field is set as 0.1MPa and the temperature is 300K. Then a treatment is made on a specific small region at [0, 5mm]×[20mm, 25mm] by setting the pressure as 2MPa and temperature as 3000K. Such procedures ignite the reactants and generates a detonation wave in one direction.

3.2. Brief discussions on numerical results.

After running for about 5 circulations, a quasi-stable detonation wave is formed, and parametric study is made thereafter. First, typical detonation characteristics ($U_D$, $P_{CJ}$ and $T_{CJ}$) are derived and compared with theoretical results after averaging over six periods. Especially $U_D$ is acquired from the circumferential denotation velocity $V_D$ and axial velocity $V_{\text{inf}}$ of injections ahead of the detonation as $U_D = \sqrt{V_D^2 + V_{inj}^2}$. Considering $V_{\text{inf}}$ keeps increasing when the position related with the velocity is away from the headwall, an average value of 370 *m/s* is chosen for representative. The theoretical values are derived by choosing the state on a typical point ahead of the detonation and using *chemkin*, and corresponding results are shown in Table 2. It can be seen that an overall error of less than 10% is acquired, and the error becomes smaller as the grid is refined. In order to check the mass conservation, the specific mass flux ratio $\frac{\dot{m}_{in}(t)}{\dot{m}_{out}(t)}$ is evaluated, where $\dot{m}(t) = \frac{1}{L} \int \rho u \cdot dl$ with *L* being the circumferential length at the inlet, and the subscripts "in" and "out" denote the injection headwall and exhaust boundary respectively. Although the histories of $\dot{m}_{in}$ and $\dot{m}_{out}$ show an oscillatory manner, the ratio of mass flux has a value oscillating around one. In Table 2, the time-averaged ratio is evaluated, where $\overline{(\cdot)}$ denotes the averaging operation. The results show values being close to one and indicate the mass conservation. Furthermore, some typical performances are evaluated in the table, e.g., the thrust $F$ and the fuel-based specific impulse $I_{sp}$. The definition of $F$ is: $F(t) = \int [\rho u^2 + (P - P_e)] dl$ and the one for $I_{sp}$ is: $I_{sp} = \frac{F(t)}{g \cdot [\int (\rho_{fuel} \cdot u) dl]}$ where $g$ is the gravity acceleration. Considering the unsteady nature of rotating detonation, the averaged values are calculated. To study the fluctuation of $I_{sp}$, a quantity is computed as $\frac{\overline{|I_{sp} - \bar{I}_{sp}|}}{\bar{I}_{sp}}$, and the small value of the quantity indicates a stable run of the detonation.

Table 2. Computational characteristics and performances in 2-D rotating detonation

|  | $U_D (m/s)$ | $P_{CJ}$ (atm) | $T_{CJ}$ (K) | $\overline{\frac{\dot{m}_{in}}{\dot{m}_{out}}}$ | $\bar{F}$ (N) | $\bar{I}_{sp}$ (s) | $\frac{\overline{|I_{sp} - \bar{I}_{sp}|}}{\bar{I}_{sp}}$ |
|---|---|---|---|---|---|---|---|
| Case 1 | 1887.29 | 30.637 | 2756 | 0.9954 | 4849.10 | 5197.88 | 1.66% |
| Case 2 | 1887.04 | 31.088 | 2801 | 0.9996 | 4964.29 | 5145.31 | 2.41% |
| Case 3 | 1891.03 | 32.087 | 2851 | 1.0022 | 5015.37 | 5125.24 | 3.49% |
| Theoret. Val. | 1996.4 | 33.942 | 3003.2 | -- | -- | -- | -- |

To explore the features of the detonation, instantaneous distributions of variables are checked in Fig.6-7 at the inlet and exit boundaries, where the result of Case 2 is chosen for illustration. In Fig. 5, the pressure spike at inlet is clearly resolved, which denotes the onset of the detonation. Using $P_0$ as a reference, it can be seen that at the inlet range ($x$=1.9~5.3cm) the pressure is above $P_0$, which is consistent to the blockage of the fuel injection. At the range ($x$=4.1~7.8cm) on exit, the pressure has a distribution approximately along $P=P_e$, which implies the subsonic exhaustion therein (see boundary condition in Section 2.3). In Fig. 6, a non-zero distribution of $v$ velocity at inlet exists at about $x$=1.9~5.3cm, which implies the blocking state after detonation as shown in Fig. 5. What is more, the existence of negative distribution indicates the continuous expansion after the detonation, which is consistent to the result in Ref. [21]. In addition, a velocity discontinuity at $x$≈4.1cm on the exit boundary is observed, which indicates the intersection of the oblique shock with the boundary. It can be seen that the flow before the shock wave has a negative $v$ velocity and the one after the shock has a positive velocity, which implies the flow moves toward the shock at the exit. The irregular distribution after the shock implies the complex waves exist in the flow field.

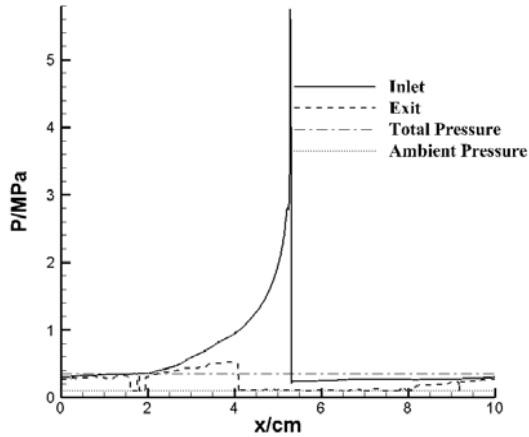 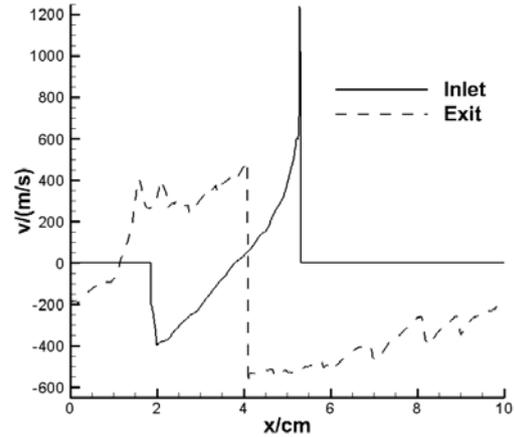

Fig. 5. Instantaneous pressure distribution at inlet and exit boundary of Case 2.

Fig. 6. Distributions of instant velocity $v$ at inlet (line) and exit (dashed line) boundary of Case 2.

In short, current computations yield reasonable results which can serve for further analysis.

## 4. Instability at interface between fresh reactants and products from previous cycle.

In previous literatures, interests regarding flow instability mainly concerned about the one of the contact discontinuity, which is originated from the inflection point where the detonation wave intersects with the oblique shock wave. In this section, the flow instability at the upstream interface between fresh injections and products from previous cycle is studied. In Ref. [14], Hishida et al. first showed the ripple structures existing on the interface (abbreviated as interface next), which was thought to be generated from Kelvin-Helmholtz instability. In order to get in-depth understanding about the mechanism, computations using high-order schemes are carried out in three grids with refining length-scales.

In Fig. 7, instantaneous temperature contours at three grids are shown at certain moments when detonation waves move at approximately the same location. In Case 1 with the coarsest grid, the computation yields a smooth description about the interface (see Fig.7(a) and Fig. 8), which is consistent to results in most literatures. The quasi-triangular zone of low temperature indicates a region of high density, and therefore a distinct feature of contact discontinuity is manifested at the interface. In Fig. 7(b), rippled structures are resolved with the grid refinement, which is quite similar to that by Hishida et al. [14]. Especially, there is obvious intrusion of the burned gas into the reactants, which might extrude the neighbor combustibles to form an "unburned gas pocket" [14]. To further explore the instability structure, a further refined grid is chosen in Case 3, and the result is shown in Fig. 7(c). It is definite that typical vortex-like roll-ups arise at the interface, and the size of the structure appears irregularly, i.e., relatively larger structures appear accompanying with the several smaller ones. What is more, intrusions in the form of slim tails are only observed from reactants into products, while the counterpart structure from products to reactants is absent. Such phenomenon indicates the role of the reactant and product is not equal in flow instability.

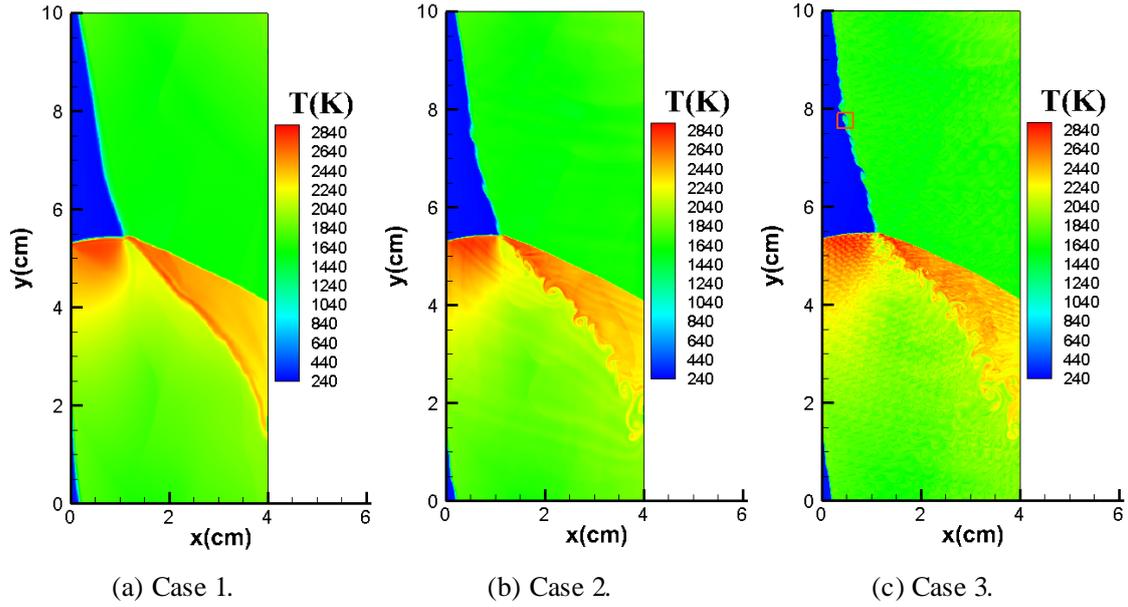

(a) Case 1.  (b) Case 2.  (c) Case 3.

Fig. 7. Instantaneous temperature contours of three cases.

Based on above observations, it is implied that there be extra complexities in the flow-field, which are previously less concerned and might contribute to flow instability. After carefully checking, two additional mechanisms are proposed besides Kelvin-Helmholtz instability [14], i.e., the effect of baroclinic torque and Rayleigh-Taylor instability. Detailed analyses are made subsequently.

(1) Kelvin-Helmholtz instability.

In this part, an analysis similar to Ref. [14] is repeated but with more detail. Because of the relatively less grids in Case 1, the flow there appears more stable and corresponding results can be treated as the base flow for stability analysis. Before further discussion, a transformation is first invoked by subtracting the velocity field with the circumferential speed of detonation wave, i.e., 1850.41m/s, so that the detonation wave could keep stationary theoretically. Then a coordinate system is configured for parametric study, where the origin is located at the joint point of detonation and oblique shock and $x'$-axis lies along the interface. Correspondingly, $u'$ and $v'$ denote velocities in $x'$ and $y'$-direction. The schematic of the system is shown in Fig. 8. Under the new system and velocity transformation, it is found that the flow moves in a direction along $x'$-axis or the interface, and different velocities exist on different sides. Hence, a typical shear layer is generated with nearly parallel flow in each side. In order to check the flow quantitatively, two points in Fig. 8 are chosen as $P_a$ and $P_b$ at (1.02434,0) and (0.499, 0), and vertical distributions of velocity $u'$ and density $\rho$ through two points are drawn in Fig. 9. From the figure, it can be seen that distributions show approximately shear-layer profiles. If the state below the $x'$ axis is denoted as "1" and the one over it is denoted as "2", then typical flow characteristics can be acquired from Fig. 9 and summarized in Table 3. Using the formula of convective velocity by Papamoschou and Roshko [19], i.e., $U_c = (\rho_1 U'_1 + \rho_2 U'_2)/(\rho_1 + \rho_2)$, the averaged speed can be evaluated as $U_c$=1980.45m/s. Because $U_c$ is thought to approximately represent the movement of the large scale vortex [19], it can also be derived from numerical result for comparison. In this regard, the result in Case 3 is chosen because the vortex-like structure there is rather discernible and easy to track. The initial location of the vortex for tracking is visualized by a box in Fig. 7(c), then $U_c$ is numerically derived as 2000m/s. Compared with the prediction, it can be seen that the difference is small. Therefore, as referred by Ref. [14], the Kelvin-Helmholtz instability should be one of the mechanisms contributing to the flow instability.

To visualize the structure of unstable structures, a further transformation is imposed on the velocity field ($u'$, $v'$) by subtracting the value $U_c$. Choosing the typical moment as that in Fig. 7(c), the instantaneous local streamlines and zoomed-in temperature contours are drawn in Fig. 10, and the region of streamlines is indicated by a box in Fig. 10(b). In Fig. 10(a), distinct vortex structures appear, and their convection effect is supposed to induce wavy structures of the interface, which is consistent to that appeared in Fig. 10(b). It is well-known that the temperature is usually used as an indicator in passive scalar study in turbulent flow to show the evolution of the interface, and in current situation the practice also favors the choice of which as the visualization variable.

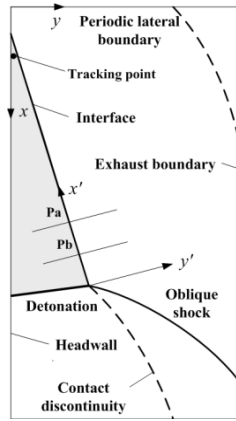
Fig. 8. Schematic of new coordinate system for analysis.

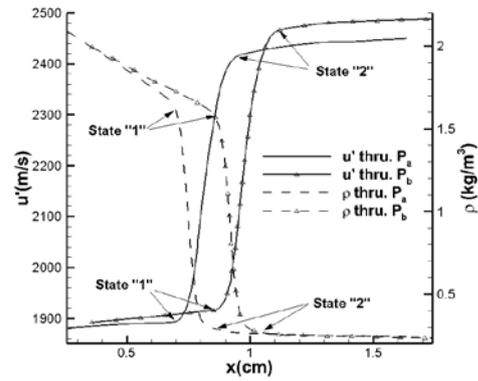
Fig. 9. Vertical distributions of velocity $u'$ and density through (abbreviated as thru.) $P_a$ and $P_b$ of Case 1.

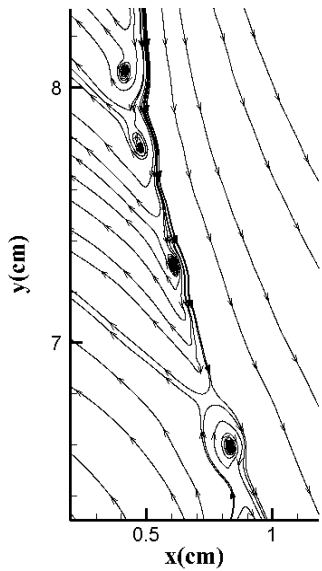
(a) Local streamlines

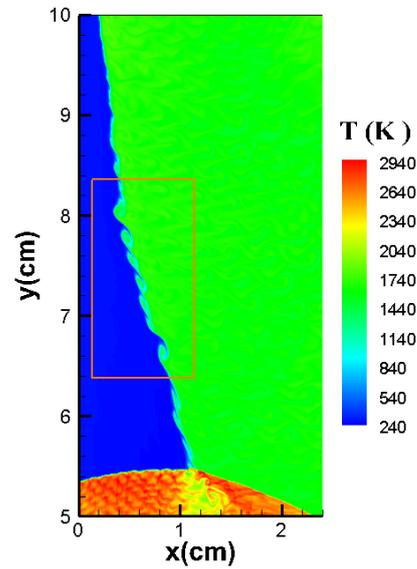
(b) Temperature contours

Fig. 10. Vortex structure at the interface shown by local streamlines and temperature contours of Case 3.

Table 3. Flow characteristics on points at two lines through $P_a$ and $P_b$ and predicted convective velocity

|  | $\rho_1(kg/m^3)$ | $U'_1(m/s)$ | $\rho_2(kg/m^3)$ | $U'_2(m/s)$ | $U_c(m/s)$ |
|---|---|---|---|---|---|
| Line through $P_a$ | 1.60331 | 1893.60 | 0.26527 | 2414.80 | 1967.59 |
| Line through $P_b$ | 1.56119 | 1916.57 | 0.25347 | 2465.88 | 1993.30 |

(2) Effect of Baroclinic torque.

In ordinary mixing layers, the pressure are usually consistent in the normal direction of the shear layer although the density might be different on different sides of the layer. In such situations the base-flow profile at least should not generate strong baroclinic torque. In the scenario of rotating detonation, the reactants are injected from high-pressured manifolds and subjected to continuous expansion until they meet products from previous cycle with relatively smaller density. Hence both gradients of pressure and density coexist at the interface and the effect of baroclinic torque might occur if their directions do not coincide with each other. To investigate this possibility, the distribution of dimensional $\left(\frac{1}{\rho^2}\nabla\rho \times \nabla P\right)^*$ is evaluated by using results of Case 1. The reason of choosing Case 1 is that the flow there is relatively stable and the attributes of the base flow can be easily obtained. After post-processing on results, dimensional baroclinic torque contours are shown in Fig. 11. From the figure, it can be seen that distinct concentrations with negative value are observed lying along the interface, and approximate zero-valued distributions exist in the region of injected reactant and burned products away from the interface. Consider the inviscid governing equation of vorticity

$$\frac{d\vec{\omega}}{dt} = (\vec{\omega} \cdot \nabla)\vec{V} - \vec{\omega}(\nabla \cdot \vec{V}) + \frac{1}{\rho^2}\nabla\rho \times \nabla P,$$

where $\vec{\omega}$ is the vorticity vector. It is expected the baroclinic torque will contribute to the evolution of vorticity. In current situation, the coherent negative vorticity exists in the interface initially, then the negative $\left(\frac{1}{\rho^2}\nabla\rho \times \nabla P\right)$ is supposed to further enforce the vorticity magnitude. This supposition can be checked by numerically evaluating the vorticity flux. As in Fig. 8, several lines perpendicular to the interface are chosen, and the vorticity flux is evaluated by $\int_l \omega^* dl$, where "$l$" denotes the line for integration. Because the vorticity mainly exists in the interface layer, the contribution of integral away from the layer will be trivial. The positions of evaluation are shown in Fig. 12(a) and the derived vorticity flux is shown in Fig. 12(b), where the abscissa is the distance measured along the fuel interface to the detonation/oblique-shock conjunction. It is definite that the vorticity flux shows an overall increase along the shear layer, which indicates the existence of baroclinic torque. With the increase of the vorticity, the shear strength is enforced, and the interface instability is potentially increased.

It is worth mentioning that there should be deflagrations along the interface which might influence the vorticity generation as well. Because simulations in this study are still not advanced enough, such effect is less quantified and will not be discussed currently.

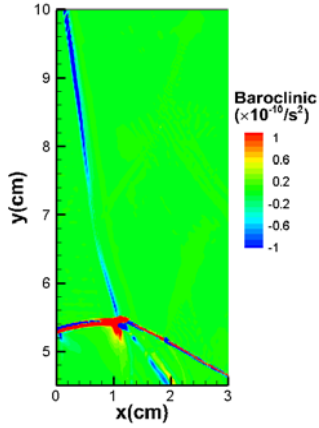 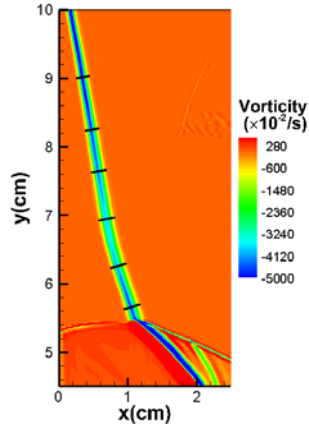 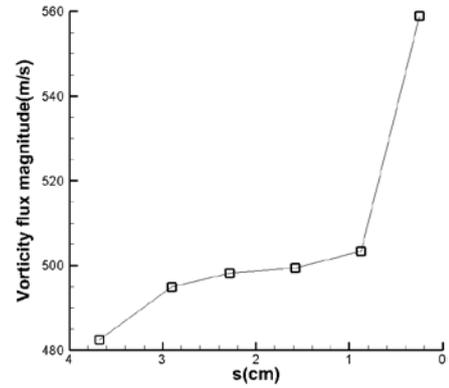

Fig. 11. Contours of $\left(\frac{1}{\rho^2}\nabla\rho \times \nabla P\right)^*$ in Case 1.

(a) Locations for investigation.

(b) Distribution of vorticity flux Magnitude.

Fig 12. Distributions of vorticity flux along fuel interface in Case 1

(3) Rayleigh-Taylor instability.

It is known from above discussions that the density of injected combustibles is different from burnt products. In rotating detonation, on the one hand the burnt gases will flow downstream with expansion, on the other hand in *x*-direction and away from the detonation, the burnt gas will endure strong expansion. Therefore the density of injected reactants is overall larger than adjacent products, which has been shown in Fig. 9. In hydrodynamics, if a heavy fluid flows toward light one with acceleration, Rayleigh-Taylor (R-T) instability might happen. During the process, some typical structures are known as [20]: spikes of heavy fluid will extrude into the light medium, while bubbles will be generated from the light fluid towards the heavy one. The structures are sketched in Fig. 13 from the courtesy of Ref. [20]. Recalling the results in Fig. 10, although not appeared in canonical symmetric forms, spike- and bubble-like structures all emerge. The phenomenon highly indicates the existence of R-T instability, and therefore it is necessary to check the occurrence conditions of the instability in rotating detonation.

First the Atwood number of the base flow is numerically derived along the interface by using the result of Case 1. From Fig. 9, the density distributes not uniformly in the upstream and downstream of the interface, but the location marking the region of discontinuity can be defined (see locations pointed by arrows in the figure). Hence the density at two places can be probed from the distribution and the Atwood number is evaluated thereby by using $(\rho_1 - \rho_2)/(\rho_2 + \rho_1)$. Following the same positions as that in Fig. 12(a), the distribution of Atwood number is acquired and drawn in Fig. 14. From the figure, Atwood numbers fall in a scope of [0.6~0.72], which are suitable for the occurrence of R-T instability.

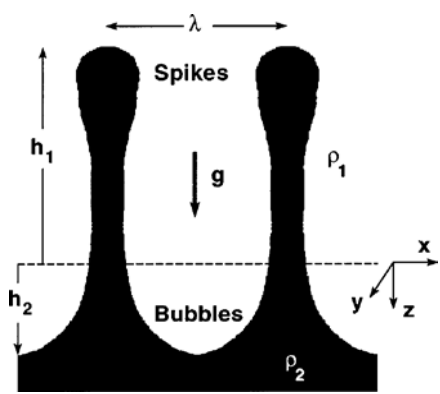 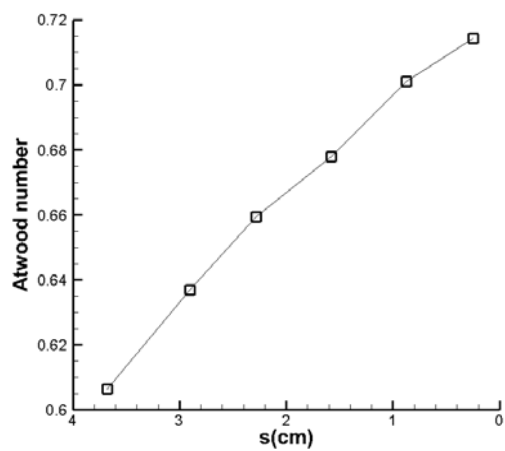

Fig. 13. Schematic of spikes and bubbles from the courtesy of Ref. [20]

Fig. 14. Distribution of Atwood number along fuel interface.

Next the movement of the interface is checked to see if it runs in acceleration by using the result of Case 2. To carry out the analysis, an initial position is chosen just after the start of injection and beneath the interface, which is indicated by "Tacking point" in Fig. 8. The same velocity transformation and corresponding framework as that in "(1) Kelvin-Helmholtz instability." is used for analysis. Based on the local velocity $(\vec{U} - V_D \vec{j})$ at the tracking point where $\vec{j}$ is the unit vector of y-axis, the new position after a time increment can be obtained by $d\vec{r}/dt = \vec{U} - V_D \vec{j}$. Recursively, the trace of the tracking point can be acquired, which approximately distributes below and along the initial instant interface as shown in Fig. 15(a). In the meanwhile, the velocity information along the trace is obtained thereafter. The history of the velocity is shown in Fig. 15(b). Because the initial tracking point is very close to the interface, the velocity of the point is supposed to approximately represent the movement of the interface (see Fig. 15(a)). From the figure, the interface shows an accelerating movement until it impinges on the detonation wave at $t \approx 315\mu s$. In detail, at $282\mu s \leq t \leq 299\mu s$, the interface moves in acceleration about $18\times10^6$ m/s$^2$; afterwards at $299\mu s < t \leq 315\mu s$, the acceleration velocity slows down to about $5.86\times10^6$ m/s$^2$. Therefore according to the computation, the interface moves in a relative stable accelerating manner from heavy reactants to light products. Similar outcomes are acquired by using results of Case 1 and 3. Hence, the two important occurrence conditions of R-T instability are satisfied. Recalling the appearance of spike and bubble-like structures, the occurrence of R-T instability is indicated.

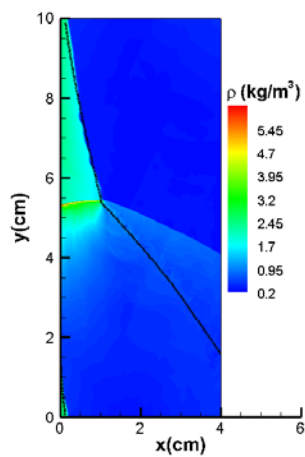 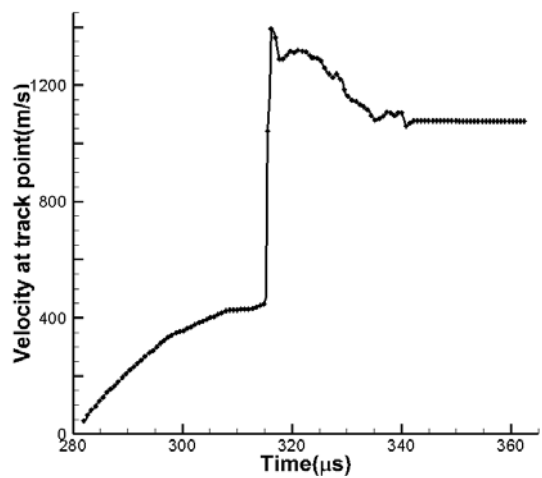

(a) Trace of point with background of initial instant density contour.

(b) History of velocity at track point.

Fig. 15. Trace and velocity history of a tracking point near the interface.

In summary, three kinds of mechanisms are investigated which should contribute to instabilities of rotating detonation, and especially evidences are provided for the two different from K-H instability. Three mechanisms are supposed to integrate together and generate specific structures along the interface.

## 5. Conclusions

Using the fifth-order WENO-PPM5 scheme to solve Euler equations with 7-species-and-8-reaction chemical model, the interface instability of a 2-D rotating detonation is further numerically studied. Three sets of grids are designed and results with different resolutions are obtained. Besides the already mentioned K-H instability, two other mechanisms are proposed, i.e., the effect of baroclinic torque and R-T instability. Brief conclusions are drawn as following:

(1) The velocity of the vortex movement at the interface is numerically derived, which is very close to the prediction by the formula of convective velocity. The quantitative coincidence confirms the validity of K-H instability as the important mechanism to generate the interface instability.

(2) The contribution of baroclinic torque with distinct magnitude is found at the interface, and the increase of vorticity magnitude is caused thereafter. The analysis indicates that the effect of baroclinic torque should increase the shear strength and the potential of instability.

(3) Separated by the interface, the heavy fresh reactants is found to inject toward the light burnt product in acceleration. Besides, typical spike- and bubble-like structures are found at the interface, which strongly suggests the occurrence of R-T instability.

**Acknowledgements**

This work was sponsored by the National Science Foundation of China under the Grant number 91541105, and also partially supported by National Key Basic Research and Development 973 Program of China under Grant Number 2014CB744100. The first author is thankful to the discussion with Prof. Frank Lu.

**Appendix**

In this study, the 7-species-and-8-reaction model regarding $H_2/O_2$ mixture by Evans & Schexnayder is used, where components are: $H_2$, $O_2$, $H$, $O$, $OH$, $H_2O$, $N_2$. The coefficients in Eq. (4) and Eq. (21) of species are given in Table A.1, which is taken from the Janef table. In the table, the absent term $A_{i,j}$ in Eq.(4) can be evaluated by $A_{i,j} = A_{i,j}^{C_p}/j$ for $j$=1…5. For each species, there are two rows of $A_{i,j}^{C_p}$ or $A_{i,j}$, where the first row corresponds to the temperature range 300K~1000K and the second row corresponds to the range 1000K~6000K.

Table A.1 Coefficients in polynomial fitting of isobaric heat capacity and specific enthalpy

| Species | $A_{i,1}^{C_p}$ | $A_{i,2}^{C_p}$ | $A_{i,3}^{C_p}$ | $A_{i,4}^{C_p}$ | $A_{i,5}^{C_p}$ | $A_{i,0}$ |
|---|---|---|---|---|---|---|
| H | 0.25000000E01 | 0.00000000E00 | 0.00000000E00 | 0.00000000E00 | 0.00000000E00 | 0.25471629E05 |
|   | 0.25000000E01 | 0.00000000E00 | 0.00000000E00 | 0.00000000E00 | 0.00000000E00 | 0.25471629E05 |
| O | 0.29464283E01 | -0.16381666E-02 | 0.24210312E-05 | -0.16028432E-08 | 0.38906964E-12 | 0.29147645E05 |
|   | 0.25420599E01 | -0.27550617E-04 | -0.31028033E-08 | 0.45510670E-11 | -0.43680515E-15 | 0.29230805E05 |
| OH | 0.38375940E01 | -0.10778857E-02 | 0.96830353E-06 | 0.18713972E-09 | -0.22571094E-12 | 0.36412822E04 |
|   | 0.29106426E01 | 0.95931650E-03 | -0.19441700E-06 | 0.13756646E-10 | 0.14224542E-15 | 0.39353816E04 |
| $H_2$ | 0.30574455E01 | 0.26765200E-02 | -0.58099158E-05 | 0.55210378E-08 | -0.18122743E-11 | -0.98890478E03 |
|   | 0.31001902E01 | 0.51119457E-03 | 0.52644211E-07 | -0.34909978E-10 | 0.36945341E-14 | -0.87738037E03 |
| $H_2O$ | 0.40701275E01 | -0.11084499E-02 | 0.41521180E-05 | -0.29637404E-08 | 0.80702101E-12 | -0.30279723E05 |
|   | 0.27167635E01 | 0.29451374E-02 | -0.80224373E-06 | 0.10226682E-09 | -0.48472138E-14 | -0.29905824E05 |
| $N_2$ | 0.36748257E01 | -0.12081501E-02 | 0.23240100E-05 | -0.63217565E-09 | -0.22577253E-12 | -0.10611587E04 |
|   | 0.28963194E01 | 0.15154865E-02 | -0.57235275E-06 | 0.99807398E-10 | -0.65223570E-14 | -0.90586181E03 |
| $O_2$ | 0.36255989E01 | -0.18782185E-02 | 0.70554543E-05 | -0.67635142E-08 | 0.21555995E-11 | -0.10475227E04 |
|   | 0.36219540E01 | 0.73618255E-03 | -0.19652231E-06 | 0.36201556E-10 | -0.28945627E-14 | -0.12019824E04 |

The reaction constants in Eq. (8) together with reactions are shown in Table A.2, where $A_j$ is given with the unit $m^3/(mol \times s)$.

Table A.2 Constants of reaction ratio in Arrhenius' law

| Reactions | Forward reaction | | | Inverse reaction | | |
|---|---|---|---|---|---|---|
|   | $A_j$ | $B_j$ | $C_j$ | $A_{-j}$ | $B_{-j}$ | $C_{-j}$ |
| $H+O_2 \Leftrightarrow OH+O$ | $2.2 \times 10^8$ | 0 | 8455 | $1.5 \times 10^{13}$ | 0 | 0 |
| $O+H_2 \Leftrightarrow OH+H$ | $7.5 \times 10^7$ | 0 | 5586 | $3.0 \times 10^{13}$ | 0 | 4429 |
| $H_2+OH \Leftrightarrow H+H_2O$ | $2.0 \times 10^7$ | 0 | 2600 | $8.4 \times 10^{13}$ | 0 | 10116 |
| $2OH \Leftrightarrow O+H_2O$ | $5.3 \times 10^6$ | 0 | 503 | $5.8 \times 10^{13}$ | 0 | 9095 |
| $H_2+X \Leftrightarrow 2H+X$ | $5.5 \times 10^{12}$ | -1 | 51987 | $1.8 \times 10^{18}$ | -1 | 0 |
| $H_2O+X \Leftrightarrow OH+H+X$ | $5.2 \times 10^{15}$ | -1.5 | 59386 | $4.4 \times 10^{20}$ | -1.5 | 0 |
| $OH+X \Leftrightarrow O+H+X$ | $8.5 \times 10^{12}$ | -1.0 | 50830 | $7.1 \times 10^{18}$ | -1.0 | 0 |
| $O_2+X \Leftrightarrow 2O+X$ | $7.2 \times 10^{12}$ | -1.0 | 59340 | $4.0 \times 10^{17}$ | -1.0 | 0 |